\documentclass[aps,twocolumn,superscriptaddress,amsmath,amsfonts,showkeys,email,
               floatfix]{revtex4}

\usepackage{epsfig,psfrag}
\usepackage{CJK}

\usepackage{graphicx}
\usepackage{dcolumn}
\usepackage{bm}
\usepackage{amssymb}
\usepackage{natbib}

\begin{document}
\title{Inelastic electron transfer in olfaction: multiphonons processes}
\author{Shu-Quan Zhang}
\affiliation{Integrated Chinese and Western Medicine Hospital, Tianjin University, Tianjin 300354 China}
\author{Yu Cui}
\affiliation{Tianjin Key Laboratory of Low Dimensional Materials Physics and Preparing Technology, Department of Physics, School of Science, Tianjin University, Tianjin 300354 China}
\author{Xue-Wei Li}
\affiliation{Tianjin Key Laboratory of Low Dimensional Materials Physics and Preparing Technology, Department of Physics, School of Science, Tianjin University, Tianjin 300354 China}
\author{Yong Sun}
\affiliation{Tianjin Key Laboratory of Low Dimensional Materials Physics and Preparing Technology, Department of Physics, School of Science, Tianjin University, Tianjin 300354 China}
\author{Zi-Wu Wang*}
\affiliation{Tianjin Key Laboratory of Low Dimensional Materials Physics and Preparing Technology, Department of Physics, School of Science, Tianjin University, Tianjin 300354 China}
\email{wangziwu@tju.edu.cn}

\begin{abstract}
Inelastic electron transfer being regarded as one of the potential mechanisms to explain the odorant recognition in the atomic-scale processes is still a matter of intense debate. Here, we propose multiphonon processes of electrons transfer using the Markvart' s model and calculate their lifetimes with values of key parameters widely adopted in olfactory systems. We find that these multiphonon processes are as quickly as the single phonon process, which suggest that contributions from different phonon modes of odorant molecule for electrons transfer in olfaction should be included. Meanwhile, temperature dependence of electron transfer could be analyzed effectively based on the reorganization energy is expanded into the linewidth of multiphonon processes. Our theoretical results not only enrich the knowledge of the mechanism of the olfaction recognition, but also provide insights for quantum processes in the biological system.
\end{abstract}
\maketitle

Our sense of smell (olfaction) is of important to make judgement about surrounding environment and influence our perceptions. In spit of the processes of olfactory recognition have been extensively studied both in experimental\cite{e1,e2,e3,j4,e4,e5,e6} and theoretical\cite{e2,e3,e6,t1,t2,t3,t4} aspects during the past several decades, the fundamental mechanism of olfaction is still hot debated. Until now, two mainstream mechanisms have been proposed: one is the docking model (lock-and-key model), in which the odorant size, shape and functional groups determine activation of olfactory receptors; another is the vibrational model, in which the special vibrational frequencies of odorant compounds contribute to odorant recognition. In particular, the latter one accounts for the essential quantum processes of electron transfer in atomic-scale level for biology systems, arousing more and more attention in recent years.
\begin{figure}
\includegraphics[width=3.4in,keepaspectratio]{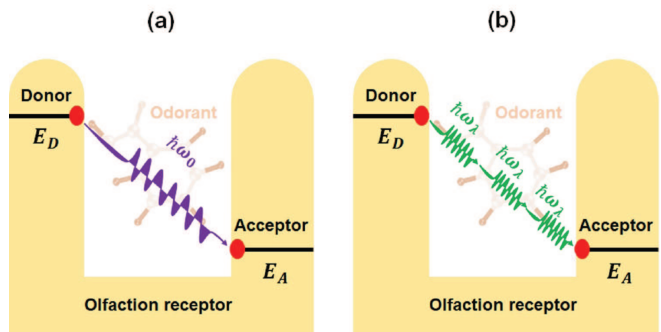}
\caption{\label{compare} The schematic diagrams of inelastic electron transfer in the olfactory receptor: (a) an electron transfers from the donor state ($E_D$) to acceptor state ($E_A$) with the absorption of a phonon ($\hbar\omega_0$) provided by the odorant molecule; (2) an electron transfers between the donor state and acceptor state by the absorption of multiphonons processes ($n\hbar\omega_{\lambda}$, $n\geq2$), where the relation of $E_D-E_A=n\hbar\omega_{\lambda}$ is satisfied.}
\end{figure}

The vibrational model original proposed first by Dyson in 1938\cite{j1} and then further studied by Wright in 1977\cite{j2}. They emphasized the contributions of these special vibrational frequencies of odorant compounds to the odor recognition. This model was endorsed by Turin in 1996\cite{j3} and proposed a quantum description that inelastic transfer of an electron mediated by an odorant in olfactory receptor. This process is schematically shown in Fig. 1 (a), in which an electron can transfer between the electron donor site ($E_D$) to the acceptor site ($E_A$)  of the olfactory receptor, but only if the energy difference ($E_{DA}$=$E_D$-$E_A$) is matched by the right vibrational energy (frequency) of the odorant molecule, and hence an odorant is recognized by its vibrational modes. Until 2007, Brookes et al\cite{t1} analyzed the physical viability of this model based on Marcus's electron transfer theory again. They evaluated the lifetime of electron transfer with the appropriate parameters of the biomolecular systems and pointed out that the time scale is sufficient to permit odorant recognition by electron transfers assisted by single vibrational modes. In recent years, this model has been further explored by a number of studies both in theory and experiments\cite{zw11,zw12,zw13,zw14,zw15,zw16}. However, only the single vibrational mode of the odorant molecular was taken into account in this model for most of these studies, the contribution of multiple vibrational modes to inelastic electron transfer are neglected.

In this paper, we theoretically propose inelastic electron transfer assisted by multiple phonons processes in olfactory receptors based on Markvart' s model\cite{zw17,zw18,zw19}. We present the lifetimes of multiphonon processes for different phonon modes and give the comparisons among them with different Huang-Rhys factors that denote the strength of electron-phonon coupling\cite{zw20}. We find that these multiphonon processes are enhanced obviously with the increasing of Huang-Rhys factor, which results in multiphonon processes are as quickly as the single phonon process. Therefore, multiphonon processes related to different phonon modes of odorant molecule should be taken into account for electrons transfer in odorant recognition. Moreover, we show temperature dependences of these multiphonon processes, in which the reorganization energy, arising from the coupling of the olfactory receptor donor and accepter states to the vibrations of the molecular environment, is expanded into the linewidth of multiphonon processes. We hope these theoretical results can trigger more experiments for the odorant recognition in future.

The model of Markvart' s multiphonon processes has been widely adopted to study the charge and energy transfer among localized states in semiconductors\cite{zw17,zw18,zw19}, in particular, to discuss the temperature dependence of multiphonon processes. We propose the inelastic electron transfer in olfactory receptor via multiphonon processes as schemed in Fig. 1 (b) and the transition rate ($1/\tau$) in this model can be expressed as
\begin{eqnarray}
\tau^{-1}&=&\frac{\sqrt{2\pi}|\gamma|^2\Omega}{\hbar[\sqrt{S_{LA}\overline{(\hbar\omega_{LA})^2_T}}+p\hbar\omega_{\alpha}]}\nonumber\\
&&\exp[-\frac{(E_{DA}-p\hbar\omega_{\alpha}-S_{LA}\overline{\hbar\omega_{LA}}}{2S_{LA}\overline{(\hbar\omega_{LA})^2_T}}],
\end{eqnarray}
with
\begin{eqnarray}
 \Omega&=&\frac{1}{\sqrt{p(1+\chi_0^2)^{1/2}}}\exp[p(\frac{\hbar\omega_{\alpha}}{2K_B T}+\sqrt{1+\chi_0^2}\nonumber\\
&&-\chi_0\cosh(\frac{\hbar\omega_{\alpha}}{2K_B T})-\ln(\frac{1+\sqrt{1+\chi_0^2}}{\chi_0}))],\nonumber\\
\end{eqnarray}
\begin{equation}
S_{LA}\overline{(\hbar\omega_{LA})^2_T}=S_{LA}\overline{(\hbar\omega_{LA})^2}\coth[\overline{\hbar\omega_{LA}}/(2K_BT)],
\end{equation}
\begin{equation}
\chi_0=\frac{S_0}{p\sinh[\hbar\omega_{\alpha}/(2K_BT)]},
\end{equation}
where $\gamma$ is the hopping integral accounting for coupling strength between the donor state and acceptor state. The reorganization energy $S_{LA}\overline{\hbar\omega_{LA}}$, arising from the the coupling of the olfactory receptor donor and accepter states to the vibrations of the molecular environment, is attributed from the acoustic phonon modes with an average phonon energy $\overline{\hbar\omega_{LA}}$ and an effective Huang-Rhys factor $S_{LA}$. In the present model, this reorganization energy is expanded into the linewidth of these multiphonon processes as given in Eq. (1), so as to the temperature dependence of multiphonon processes could be analyzed qualitatively. $p=1,2,3\cdots$ is the phonon numbers of $\hbar\omega_{\alpha}$ with the proper branches of modes $\alpha$ (frequencies) for the inelastic electron transfer. In fact, probabilities of these multiphonon processes depend on another Huang-Rhys factor $S_0$, which describes the coupling strength between the olfactory receptor and odorant molecule. $E_{DA}$ is the energy separation between the donor and acceptor states in the olfactory receptor, K$_B$ is the Bolzmann constant and $T$ is the temperature.

The characteristic values of these physical qualities above mentioned in Eq. (1) have been discussed very widely for bimolecular systems, such as the hopping integral $\gamma=1$ meV, the reorganization energy $S_{LA}\overline{\hbar\omega_{LA}}$ is expected to lies in the range of 30-50 meV\cite{t1,t3,zw12} and Huang-Rhys factor could be varied in the range of 0-0.35 due to the contribution from different phonon modes\cite{zw11}. The adopted values of these quantities for the simulations in this paper have been listed in table I.
\begin{table}[htbp]
\caption{\label{compare}These adopted parameters in the numerical simulations for multiphonon processes of inelastic electron transfer in olfactory receptor.}
\begin{tabular}{ccccccccccc}
\hline
\\
Parameter         &   Values \\[1.0ex]\hline
Hopping integral constant ($\gamma$)    &  1.0 (meV)$^{8,10}$   \\
The reorganization energy ($S_{LA}\overline{\hbar\omega_{LA}}$) &   30 (meV)$^{8,10,15}$  \\
Average phonon energy ($\overline{\hbar\omega_{LA}}$) & 6.0 (meV)  \\
Donor-acceptor energy separation ($E_{DA}$)         & 200 (meV) $^{8,10,15}$  \\
Huang-Rhys factor  ($S_0$)   &      0$\sim$0.35$^{8,15,16}$    \\
The energy of phonon mode ($\hbar\omega_{\alpha}$) &      Variable    \\
\hline
\end{tabular}
\end{table}
\begin{figure*}
\centering
\includegraphics[width=7.1in,keepaspectratio]{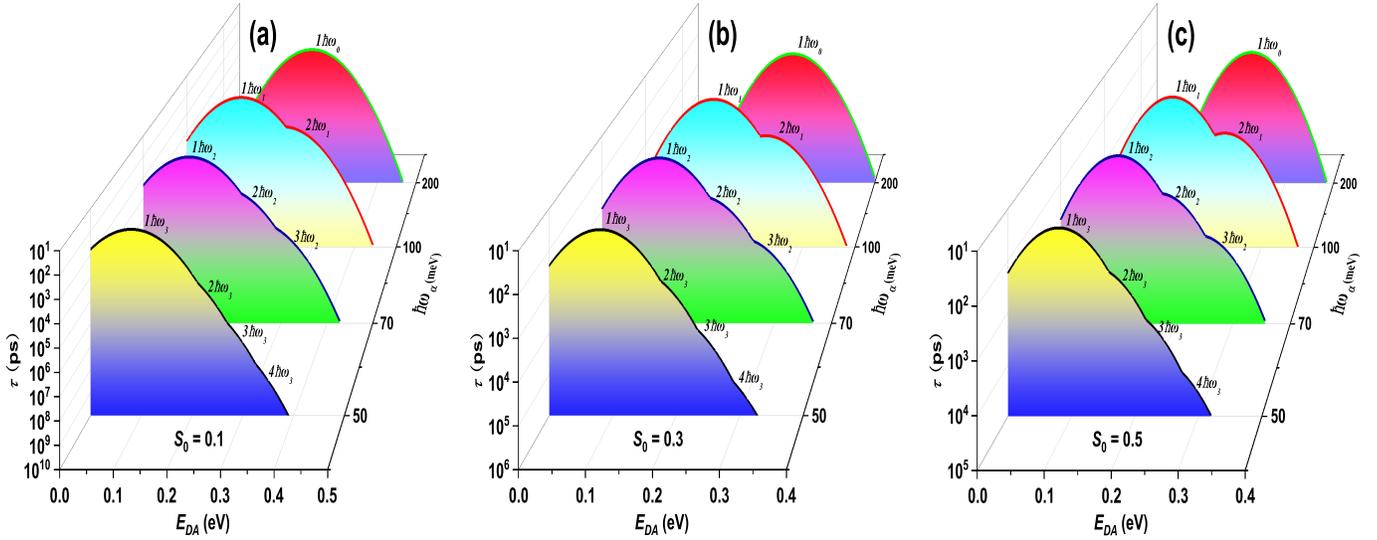}
\caption{Lifetimes of inelastic electron transfer assisted by one ($\hbar\omega_0$=200 meV) phonon, two ($\hbar\omega_1$=100 meV), three ($\hbar\omega_2$=70 meV) and four ($\hbar\omega_3$=50 meV)  phonons processes at $E_{DA}$=200 meV and $T$=309 K (body temperature). (a), (b) and (c) for Huang-Rhys factor $S_0$=0.1, 0.3, 0.5, respectively.}
\end{figure*}
\begin{figure*}
\centering
\includegraphics[width=7.5in,keepaspectratio]{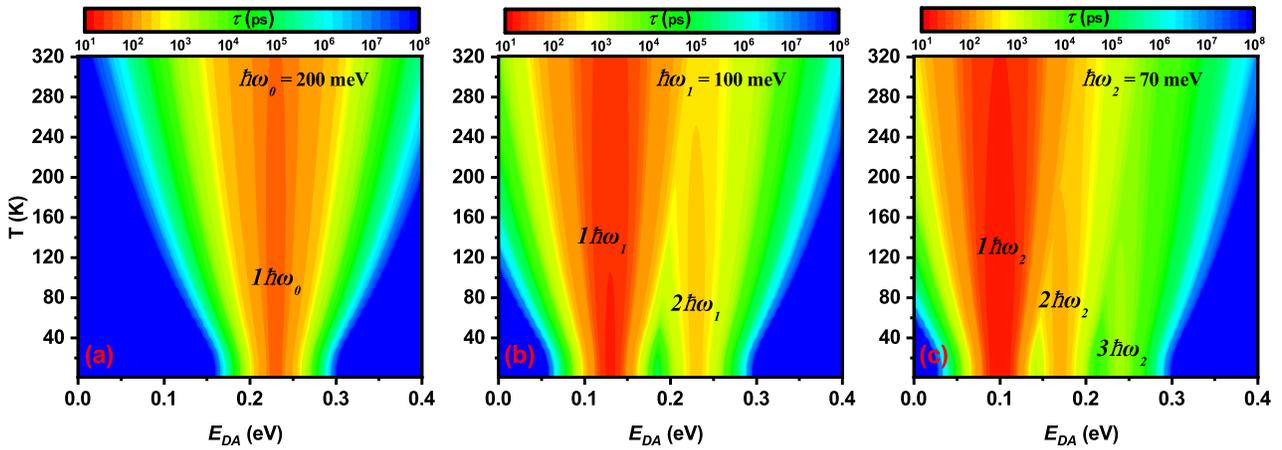}
\caption{Temperature dependence of inelastic electron transfer assisted by different phonon modes at $E_{DA}$=200 meV and $S_0$=0.3. (a), (b) and (c) for one phonon ($\hbar\omega_0$=200 meV), two ($\hbar\omega_1$=100 meV) and three ($\hbar\omega_2$=70 meV) phonons processes, respectively.}
\end{figure*}

The value of energy difference $E_{DA}$=200 meV between the donor and acceptor states in the olfactory receptor has been widely adopted by Brookes and others\cite{t1,t3,zw11,zw12}. They studied the process of an electron transfers when $E_{DA}$ is matching the energy of the single phonon based on the well-known Marcus' model. They pointed out that the corresponding lifetime of electron transfer should be much faster than the process of elastic electron tunneling, making the inelastic signal sufficiently strong relative to the elastic signal in olfactory receptor. But energies of multiple phonons matching the energy difference $E_{DA}$ have been neglected in these works even though the odorant molecule provides different phonon modes. We present multiphonon processes for four types of phonon modes $\hbar\omega_0=200$ meV,  $\hbar\omega_1=100$ meV,  $\hbar\omega_2=70$ meV and $\hbar\omega_3=50$ meV, respectively, and give the comparisons for them in Fig. 2 at different values of Huang-Rhys factor ($S_0$) .  From Fig. 2 (a), one can see that the lifetime of single phonon process is in the range of tens of picosecond given by this model, which is much faster than the predicted time scale (nanosecond) by Brookes et al\cite{t1,t3,zw11}. Moreover, two- and three-phonons processes are also in the range of nanoseconds, which are also sufficient to permit detection on the observed time scale\cite{t1}. These multiphonon processes are enhanced obviously with increasing Huang-Rhys factor $S_0$ as shown from Fig. 2 (a) to (c). Therefore, these multiphonon processes should give the significant contributions to inelastic electron transfer, in particular, for the energy difference $E_{DA}$ is much larger than the single phonon energy that the odorant molecules provided. Here, we must emphasize that Huang-Rhys factor $S_0$ is a very important parameter for these multiphonon processes and its value can varies in the range of 0.0 $\sim$ 0.35 for the typical odorants in some previous studies\cite{zw11,zw12}. However, this factor is related to many microscopic electronic structures of the olfactory receptor, which results in the accurate evaluation of Huang-Rhys factor is still a challenge work both in the theoretical and experimental aspects.  In addition, the combined multiphonon processes consisting of different phonon modes, such as $\hbar\omega_1+2\hbar\omega_3$ and $2\hbar\omega_2+\hbar\omega_3$, should also be taken into account if the larger value of Huang-Rhys factor could be obtained.

The temperature dependences of inelastic electron transfer via three types of phonon modes are shown in Fig. 3, where an effective phonon modes $\overline{\hbar\omega_{LA}}$ =6 meV is assumed for the reorganization energy. One can see that the more broader energy window is opened for electron transfer with increasing the temperature, so the influence of the thermal fluctuation of the molecular environment on the odorant recognition could be analyzed qualitatively based on the present model, which is also an advantage of this model to study the quantum mechanism of the olfaction. In fact, the similar results could be obtained when the adopted effective phonon mode $\overline{\hbar\omega_{LA}}$ is in the range of several meV for the reorganization energy. Lastly, the olfactory discrimination of enantiomers with chiral properties  based on the inelastic electron transfer mechanism has been proposed\cite{zw13}. These multiphonon processes for the chiral odorants are not considered, which will also be very important for the olfaction study and need to further investigate in the following work.

In conclusion, we theoretically study electrons transfer via multiphonon processes based on the Markvart' s model and stimulate these processes with values of key parameters usually adopted in olfactory receptors. We find that (1) these multiphonon processes are as quickly as the single phonon process, which suggest that inelastic electrons transfer in odorant recognition should include contributions from different phonon modes of odorant molecule; (2) more broader energy window is opened for electron transfer with increasing the temperature, since the influence of the thermal fluctuation of the molecular environment on the odorant recognition could be analyzed effectively with an effective phonon modes $\overline{\hbar\omega_{LA}}$ is adopted for the reorganization energy.

\section{Acknowledgement}
This work was supported by National Natural Science Foundation of China (Grant No 11674241).

\section{Data availability}
The data that support the findings of this study are available from the corresponding author upon reasonable request.

\end{document}